\documentclass[epj]{svjour}
\usepackage{latexsym}
\usepackage{amsmath}
\usepackage{graphics}

\newcommand{\beq}{\begin{equation}}
\newcommand{\eeq}{\end{equation}}    
\newcommand{\beqar}{\begin{eqnarray}}
\newcommand{\eeqar}{\end{eqnarray}} 
\newcommand{\ds}{\displaystyle} 
\newcommand{\llangle}{\left\langle\!\left\langle}
\newcommand{\rrangle}{\right\rangle\!\right\rangle}
\usepackage{color}

\begin{document}
\title{Dynamical vs geometric anisotropy in relativistic heavy-ion 
collisions: which one prevails?}
\author{L.V.~Bravina \inst{2,3} 
\thanks{e-mail: larissa.bravina@fys.uio.no}
\and I.P.~Lokhtin \inst{1}
\and L.V.~Malinina \inst{1}
\and S.V.~Petrushanko \inst{1}
\and A.M.~Snigirev \inst{1}
\and E.E.~Zabrodin \inst{1,2,3}}
\institute{
Skobeltsyn Institute of Nuclear Physics, Lomonosov Moscow State
University, Moscow, Russia \and
Department of Physics, University of Oslo, Oslo, Norway \and
National Research Nuclear University "MEPhI" (Moscow Engineering Physics
Institute), Moscow, Russia 
 }
\date{Received: date / Revised version: date}
%
\abstract{
We study the influence of geometric and dynamical anisotropies on the 
development of flow harmonics and, simultaneously, on the second- and
third-order oscillations of femtoscopy radii. The analysis is done 
within the Monte Carlo event generator HYDJET++, which was extended to
dynamical triangular deformations. It is shown that the merely geometric 
anisotropy provides the results which anticorrelate with the 
experimental observations of either $v_2$ (or $v_3$) or second-order 
(or third-order) oscillations of the femtoscopy radii. Decays of 
resonances significantly increase the emitting areas but do not change
the phases of the radii oscillations. In contrast to 
the spatial deformations, the dynamical anisotropy alone provides the 
correct qualitative description of the flow and the femtoscopy 
observables simultaneously. However, one needs both types of the 
anisotropy to match quantitatively the experimental data.   
} 
\maketitle
\section{Introduction}
\label{intro}

Search for the signals of a new state of matter, quark-gluon plasma
(QGP), is one of the main goals of experiments on heavy-ion collisions 
at (ultra)relativistic energies at modern colliders RHIC and LHC and 
at coming soon facilities FAIR and NICA. The QGP is formed during the
short highly non-equilibrium stage, when two relativistic nuclei smash
each other and produce a hot expanding fireball. The quark-gluon plasma
in the fireball quickly relaxes to thermodynamic equilibrium. Since the
plasma is not believed anymore to be a weakly interacting gas of quarks 
and gluons but rather considered to be a strongly interacting liquid 
\cite{Shur_04}, 
its further evolution is treated within the framework of relativistic
hydrodynamics \cite{Land_53,Bjor_83}. Fireball expansion leads to a 
decrease of the temperature, and at a certain moment it reaches the
temperature of quark-hadron phase transition. The QGP hadronizes, but
the fireball continues to expand until the thermal contact between the
particles is lost. This is the stage of thermal freeze-out. After the
decays of resonances, individual hadrons will hit various detectors.
The main problem for both theoreticians and experimentalists is to 
search for the QGP fingerprints on the reconstructed particle yields and
energy spectra.

Anisotropic collective flow of hadrons in non-central heavy-ion 
collisions appears to be one of the few signals extremely sensitive to
even a small amount of created quark-gluon plasma. The flow is analyzed
in terms of Fourier series expansion of particle distribution in 
azimuthal plane \cite{VoZh_96,VPS_10}


\beq
\ds
\frac{d N}{d \phi} \propto 1 + 2 \sum \limits_{n=1}^{\infty}
v_n \cos{\left[ n (\phi - \Psi_{EP,n}) \right] }~.
\label{eq1}
\eeq
Here $\phi$ is the azimuthal angle between the particle transverse 
momentum and the participant event plane, and $\Psi_{EP,n}$ denotes the 
azimuth of the event plane of $n$-th flow component, respectively. The 
Fourier coefficients $v_n$ represent the flow harmonics,
\beq
\ds
v_n = \llangle \cos{\left[ n (\phi - \Psi_{EP,n})\right] }\rrangle~.
\label{eq2}
\eeq   
The averaging in Eq.~(\ref{eq2}) is performed over all particles in a
single event and over all events. The first coefficients are dubbed
{\it directed}, $v_1$, {\it elliptic}, $v_2$, {\it triangular}, $v_3$, 
{\it quadrangular}, $v_4$, flow, and so forth. The reason of the 
anisotropic flow development in the system is the translation of the
spatial anisotropies $\varepsilon_n$ of the overlapping zone into the 
momentum anisotropies $v_n$ of final hadronic distribution. Recall that 
because of the initial state fluctuations the spatial anisotropies occur
even in very central nuclear collisions. Also, momentum anisotropy may 
arise due to the non-isotropic azimuthal dependence of the transverse 
velocity of expanding fireball which leads to the adjustment of the 
collective flow gradients in various directions. To distinguish between 
the two sources of particle momentum anisotropy we will call it 
geometric and dynamical anisotropy, respectively. These anisotropies 
should affect the two-particle femtoscopy correlations used to restore 
the size and the shape of the emitting source.

Generally, the femtoscopy correlations \cite{HBT,GGL,pod89}
are measured as a function of 
pair relative momentum four vector q. An invariant form of this
momentum difference commonly used in the one dimensional correlation
analysis is $q_{inv} = \sqrt{q_{0}^{2} - |q|^{2}} $, and the correlation 
function (CF) is represented by a single-Gaussian
\beq
CF_{single}(q_{inv}) =  1 + \lambda\exp{\left(
-R_{inv}^{2} q^{2}_{inv}\right)}~,
\label{eq:single}
\eeq
where the parameters $R_{inv}$ and $\lambda$ indicate the size of the 
emitting source and the correlation strength, respectively. Note that 
$R_{inv}$ in Eq.(\ref{eq:single}) is defined in the pair rest frame 
(PRF). 

The more advanced technique is the 3-dimensional correlation analysis.
Here the momentum and directional dependence of the correlation 
function can be used to get information about the shape of the emission 
region and the duration of the emission in order to reveal the details 
of the production dynamics \cite{pod89,led04,lis05}. In such a 3D
analysis the correlation functions are studied in terms of the $out$, 
$side$ and $longitudinal$ components of the relative momentum vector
${\bf q}=\{q_{out},q_{side},q_{long}\}$ \cite{pod83,bdh94}. Here the
longitudinal component of the vector ${\bf q}$ is parallel to the beam
axis. The orthogonal transverse components, $q_{out}$ and $q_{side}$, 
are oriented in such a way that the direction of $q_{out}$ is parallel 
to the pair transverse velocity, and $\{ q_{out},q_{side},q_{long}\} 
\equiv \{ q_o,q_s,q_l \}$ is a right-handed system. The corresponding 
widths of the CF are commonly parametrised in terms of the Gaussian 
correlation radii $R_o, R_s, R_l$ and their cross terms
\beqar \ds
 \label{eq:CF3D}
\nonumber
& & CF(\mathrm{q},\Phi) - 1 = \\
& & \lambda \exp \left[
 - R_\mathrm{o}^2(\Phi)q_\mathrm{o}^2
 - R_\mathrm{s}^2(\Phi)q_\mathrm{s}^2
 - R_\mathrm{l}^2(\Phi)q_\mathrm{l}^2 \right. \\
\nonumber  
& &\left.  - R_\mathrm{o,s}^2(\Phi)q_\mathrm{o} q_\mathrm{s} 
  - R_\mathrm{o,l}^2(\Phi)q_\mathrm{o} q_\mathrm{l} 
  - R_\mathrm{s,l}^2(\Phi)q_\mathrm{s} q_\mathrm{l} 
\right]~,
\eeqar
where $\Phi \equiv \phi_{pair}$ is the azimuthal angle of the pair 
three-momentum with respect to the reaction plane $z$-$x$ determined by 
the longitudinal direction and the direction of the impact parameter 
vector.
This 3D analysis is performed in the so-called longitudinal comoving 
system (LC\-MS), in which the pair momentum along the beam axis is zero.
In the boost-invariant case, the transverse-longitu\-di\-nal cross terms 
$\it{out,long}$ and $\it{side,long}$ vanish in the LCMS frame, 
whereas the $\it{out,side}$ cross term can be present.

The analysis is usually carried out for different collision centralities; 
and avarage transverse momentum of the pair ranges 
${\bf k}_{\mathrm T} = ({\bf p}_{\mathrm{T,1}} + {\bf p}_{\mathrm{T,2}}) 
/2$.  A more differential femtoscopic analysis is performed in bins of 
$\Delta\phi_n  = \phi_{pair} - \Psi_{EP,n}$ defined in the range 
$(0, \pi)$, where $\Psi_{EP,n}$ is the $n$-th order event-plane angle. 

In the Gaussian approximation, the radii in the Eq.(\ref{eq:CF3D}) are 
related to space-time variances via the set of equations 
\cite{Wiedemann:1997cr}:
\beqar \ds
\label{eq:radii}
\nonumber
R_\mathrm{s}^2 &=& 
\frac{\langle \widetilde{x}^2 \rangle+\langle \widetilde{y}^2\rangle}{2}
-\frac{\langle \widetilde{x}^2 \rangle-\langle \widetilde{y}^2 \rangle}
{2} \cos(2\Phi)-
\langle \widetilde{x}\widetilde{y} \rangle\sin(2\Phi)~,
\\
\nonumber
R_\mathrm{o}^2 &=&
\frac{\langle \widetilde{x}^2 \rangle+\langle \widetilde{y}^2\rangle}{2}
+\frac{\langle \widetilde{x}^2 \rangle-\langle \widetilde{y}^2 \rangle}
{2} \cos(2\Phi)
+\langle \widetilde{x}\widetilde{y} \rangle\sin(2\Phi)
\\
&-& 2\beta_{T} \left[ \langle \widetilde{t}\widetilde{x} \rangle
\cos(\Phi)+\langle \widetilde{t}\widetilde{y} \rangle\sin(\Phi)\right] + 
\beta_{T}^{2} \langle \widetilde{t}^2\rangle~,
\\
\nonumber
R_\mathrm{l}^2 &=& \langle \widetilde{z}^2 \rangle -2 \beta_l \langle 
\widetilde{t}\widetilde{z} \rangle
+ \beta_{l}^{2} \langle \widetilde{t}^2\rangle~, 
\\
\nonumber
R_\mathrm{o,s}^2 &=&
\langle \widetilde{x}\widetilde{y} \rangle\cos(2\Phi)
-1/2(\langle \widetilde{x}^2 \rangle-\langle \widetilde{y}^2 \rangle) 
\sin(2\Phi) 
\\
\nonumber
&+& \beta_{T} \left[ \langle \widetilde{t}\widetilde{x} \rangle
\sin(\Phi)- \langle \widetilde{t}\widetilde{y} \rangle\cos(\Phi) 
\right]~. 
\eeqar
Here $\beta_l=k_z/k^0$, $\beta_{T}=k_{\rm T}/k^0$, and $\Phi = 
\phi_{pair}$, respectively.
The space-time coordinates $\widetilde{x}^{\mu}$ are defined relative 
to the effective source center $\langle x^{\mu}\rangle$ as
$\widetilde{x}^{\mu}=x^{\mu}-\langle x^{\mu}\rangle $.
The averages are taken with the source emission function
$S(t, \vec{x}, k)$ \cite{Wiedemann:1997cr}
\beq \ds
\langle f(t, \vec{x}) \rangle =\frac{\int{d^4 x f(t, \vec{x})S(t, 
\vec{x},k)}} {\int{d^4 x S(t, \vec{x},k)}}~.
\eeq

The sensitivity of femtoscopy correlations alone, and together with
directed, elliptic or triangular flow, to the source ani\-so\-tro\-py 
was studied in many papers, see, e.g., 
\cite{Wiedemann:1997cr,LHW_plb00,HK_plb02,RL_prc04,CTC_epja08,PSH_13,LCTC_epja16,CTCL_17} 
and references therein. For the analysis the authors have employed an 
ideal hydrodynamic model, a toy Gaussian-source model, the Buda-Lund 
model \cite{CL_prc96}, and the blast-wave model \cite{SR_prl79}. 
The study of second-order (i.e., with respect to reaction plane of 
elliptic flow) and third-order (with respect to the reaction plane of 
triangular flow) harmonic oscillations of the femtoscopy radii has shown 
that geometric anisotropy seems to determine the second-order radii 
oscillations \cite{HK_plb02}, whereas the third-order oscillations are
dominated by the triangular flow anisotropy \cite{PSH_13}. 
The possibility of disentangling of geometric and dynamical anisotropies 
by means of the simultaneous analysis of the flow harmonics and 
femtoscopy observables becomes a very popular topic nowadays.

In our analysis of disentangling between the both ani\-so\-tro\-pies we 
apply the HYDJET++ model 
\cite{Lokhtin:2008xi,Lokhtin:2012re,Bravina:2013xla} 
which relies on parametrisation of the freeze-out state similar to the 
Buda-Lund model and the Cracow model THERMINATOR \cite{THERM}. The major 
difference between the models is that HYDJET++ treats also hard 
processes in addition to para\-me\-trised hydrodynamics. Since the model 
allows one to switch on/off both dynamical and geometric anisotropy 
parameters independently, in the present paper we investigate the 
separate influence of each factor on elliptic $v_2$ and triangular $v_3$ 
flow and, simultaneously, on the radii $R_{side},\ R_{out},\ R_{long}$ 
of the fireball. The model employs a very extensive table of resonances 
with more than 360 baryons, mesons, and their antistates. Therefore, it 
would be very tempting to examine the influence of resonance decays 
on the oscillations of the femtoscopic radii. 

The paper is organized as follows. Description of HYDJET++ is given in 
Sec.~\ref{sec:sec2}. Special attention is paid to parameters responsible 
for formation of geometric and dynamical second-order and third-order 
anisotropies of the fireball. Section~\ref{sec:sec3} presents the 
results concerning the influence of key anisotropy parameters of the 
model on both the flow harmonics and the femtoscopy radii simultaneously. 
Conclusions are drawn in Sec.~\ref{concl}.
 
\section{HYDrodynamics with JETs (HYDJET++)}
\label{sec:sec2}

HYDJET++ is a model of relativistic heavy ion collisions which 
incorporates two independent components: the soft hydro-type state and 
the hard state resulting from the medium-modified multi-parton 
fragmentation. The details of the model and corresponding simulation 
procedure can be found in the HYDJET++ manual~\cite{Lokhtin:2008xi}. Its 
input parameters have been tuned to reproduce the experimental LHC data 
on various physical observables~\cite{Lokhtin:2012re,Bravina:2013xla} 
measured in Pb+Pb collisions at center-of-mass energy 2.76 TeV per 
nucleon pair, namely, centrality and pseudorapidity dependence of 
inclusive charged particle multiplicity, transverse momentum spectra 
and $\pi^\pm \pi^\pm$ correlation radii in central Pb+Pb collisions, 
momentum and centrality dependencies of elliptic and higher-order
harmonic coefficients. The main features of the model valuable for the 
current studies are briefly listed below.

The soft component represents the hadronic state generated on the 
chemical and thermal freeze-out hypersurfaces obtained from the 
parametrisation of relativistic hydrodynamics with preset freeze-out 
conditions (the adapted event generator 
FAST MC~\cite{Amelin:2006qe,Amelin:2007ic}). It is supposed that a 
hydrodynamic expansion of the fireball ends by a sudden system breakup
(``freeze-out'') at given temperature $T$. The scenario with different 
chemical and thermal freeze-outs is implemented in HYDJET++. It means 
that particle number ratios are fixed at chemical freeze-out temperature 
$T^{\rm ch}$, while the effective thermal volume $V_{\rm eff}$ and 
hadron momentum spectra being computed at thermal freeze-out temperature 
$T^{\rm th} \le T^{\rm ch}$.

The direction and strength of the elliptic flow in the model are 
governed by two parameters. The spatial aniso\-t\-ro\-py 
$\varepsilon_2(b)$ represents the elliptic modulation of the final 
freeze-out hypersurface at a given impact parameter $b$, whereas the 
momentum anisotropy $\delta_2(b)$ deals with the modulation of flow 
velocity profile. The transverse radius of the fireball reads
\beq \ds
R_{\rm ell}(b, \phi) = R_{\rm f}(b) \frac{\sqrt{1 - \varepsilon_2^2(b)}}
{\sqrt{1 + \varepsilon_2(b) \cos{2\phi}}}~,
\label{eq:5}
\eeq
where
\beq \ds
R_{\rm f}(b) = R_0 \sqrt{1 - \varepsilon_2(b)}~.
\label{eq:6}
\eeq
In the last equation $R_0$ denotes the freeze-out transverse radius in 
case of absolutely central collision with $b = 0$. Then, the spatial 
anisotropy is transformed into the momentum anisotropy at the 
freeze-out, because each of the fluid cells is carrying a certain 
momentum. Dynamical anisotropy implies that the azimuthal angle of the 
fluid cell velocity, $\phi_{\rm cell}$, does not coincide with the 
azimuthal angle $\phi$, but rather correlates with it 
\cite{Amelin:2007ic} via the nonlinear function containing the 
anisotropy parameter $\delta_2(b)$
\beq \ds
\frac{\tan{\phi_{\rm cell}}}{\tan{\phi}} = 
\sqrt{\frac{1 - \delta_2(b)}{1 + \delta_2(b)}}~.
\label{eq:7}
\eeq
As was mentioned in \cite{Amelin:2007ic}, in case of $\delta_2 \neq 0$
even the spherically symmetric source can mimic the spatially contracted
one.
Both $\delta_2 (b)$ and $\varepsilon_2 (b)$ can be treated independently 
for each centrality, or may be related to each other through the 
dependence of the elliptic flow coefficient $v_2(\varepsilon,\delta_2)$ 
obtained in the hydrodynamical approach~\cite{Wiedemann:1997cr}:
\begin{equation}
\label{v2-eps-delta1}
v_2(\varepsilon_2, \delta_2) \propto \frac{2(\delta_2 - \varepsilon_2)}
{(1-\delta_2^2)(1-\varepsilon_2^{2})}~.
\end{equation}

To extend the model for triangular flow we have to introduce another 
parameter, $\varepsilon_{3}(b)$, which is responsible for the spatial 
triangularity of the fireball. The altered radius of the freeze-out 
hypersurface in azimuthal plane reads
\begin{equation}
\displaystyle
\label{eq:9}
R(b,\phi) =R_{\rm ell}(b)[1+\varepsilon_{3}(b)\cos{[3(\phi -
\Psi_{EP,3})]}]~.
\end{equation}
The experimental data indicate that elliptic flow does not correlate 
with triangular flow. Therefore, the event plane of the triangular flow, 
$\Psi_{EP,3}$, is randomly oriented with respect to the plane 
$\Psi_{EP,2}$, which is fixed to zero in the model, thus providing the 
independent generation of elliptic and triangular flow. Triangular 
dynamical anisotropy can be introduced, for instance, via the 
parametrisation of maximal transverse flow rapidity \cite{Lokhtin:2008xi} 
\beq \ds
\label{eq:10}
\rho_{\rm u}^{max}(b) = \rho_{\rm u}^{max}(0)\left\{ 1+ \rho_3(b) 
\cos{[3(\phi-\Psi_{EP,3})]} + \ldots \right\}~,
\eeq
where $u$ is the 4-velocity of the fluid cell. In this case we are 
getting the triangular modulation of the velocity profile on the whole
freeze-out hypersurface by introducing the new anisotropy parameter,
$\rho_3(b)$. Again, this parameter can be treated independently for each 
centrality, or can be expressed through the initial ellipticity 
$\varepsilon_0(b)=b/2R_A$, with $R_A$ being the radius of colliding 
nuclei. The particular role of each of the anisotropy parameters, 
$\varepsilon_2, \delta_2, \varepsilon_3,\ {\rm and}\ \rho_3$, in the 
formation of the flow harmonics and femtoscopy correlations is clarified 
in Sec.~\ref{sec:sec3}.

The approach for the hard component is based on the PYQUEN jet quenching
model~\cite{Lokhtin:2005px} modifying the nucleon-nucleon collisions 
generated with PYTHIA$\_$6.4 ev\-ent generator~\cite{Sjostrand:2006za}. 
The radiative partonic energy loss is computed within BDMPS
model~\cite{Baier:1996kr,Baier:1999ds,Baier:2001qw}, whereas the 
collisional energy loss due to elastic scatterings and the dominant 
contribution to the differential scattering cross section being 
calculated in the high-momentum transfer 
limit \cite{Bjorken:1982tu,Braaten:1991jj,Lokhtin:2000wm}. The effect
of nuclear shadowing on parton distribution functions is taken into 
account for hard component using the impact parameter dependent 
pa\-ra\-met\-ri\-za\-ti\-on~\cite{Tywoniuk:2007xy} obtained in the
framework of Glauber-Gribov theory.

To study the femtoscopy momentum correlations, it is necessary to 
specify a space-time structure of a hadron emission source. The 
treatment of the coordinate information for particles from soft 
component and for low momentum jet particles (with $p_{\rm T}<1$~GeV/$c$) 
in the model is similar: such particles are emitted from the fireball of 
radius $R_{\rm f}$ at mean proper time $\tau_{\rm f}$ with the emission 
duration $\Delta \tau_{\rm f}$. Similar to any Bjorken-like model with 
cylindrical parametrisation, HYDJET++ transforms the azimuthal 
anisotropy of the freeze-out hypersurface into the azimuthal anisotropy 
of the particle momentum distribution proportionally to a term 
$(p_{\rm} \sinh{Y_{\rm T}}/T^{\rm th}) \cos{(\phi -\varphi)}$ 
\cite{Lokhtin:2008xi}, arising in scalar product of 4-vectors of particle 
momentum and flow velocity of the fluid element. Here $\phi$ and 
$\varphi$ are the azimuthal angles of the fluid element and of the 
particle, respectively, and $Y_{\rm T}$ is the transverse flow rapidity. 
Four-coordinates of high momentum jet particles (with $p_{\rm T} > 
1$~GeV/$c$) are coded in a bit different way~\cite{Lokhtin:2012re}, but 
this aspect of the model is out of the current paper scope and
does not affect our present consideration.

Further details of the HYDJET++ model can be found elsewhere
\cite{Lokhtin:2008xi,Lokhtin:2012re,Bravina:2013xla}.
The model was successfully applied for the description of various
signals in ultra-relativistic heavy ion collisions, including elliptic
\cite{v2_prc09,v2_sqm09} and triangular flow \cite{v3_prc17,v3_sqm15},
higher flow harmonics up to hexagonal flow 
\cite{Bravina:2013xla,v4_prc13,v6_prc14}, azimuthal dihadron correlations 
(ridge) \cite{ridge_prc15}, event-by-event fluctuations of the flow
harmonics \cite{fluct_epjc15}, and flow of mesons with open and hidden 
charm \cite{charm_arx}.

\section{Influence of dynamical and geometric anisotropies on flow and
femtoscopy observables}
\label{sec:sec3}

\begin{figure}
\centering
\resizebox{0.45\textwidth}{!}{
\includegraphics{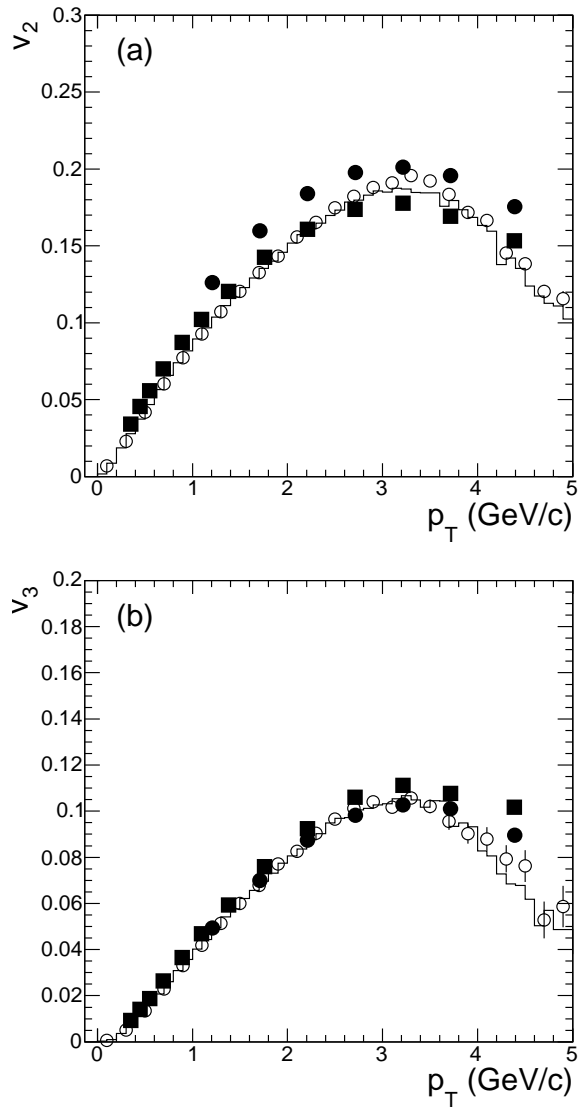}}
\caption{
(a) Elliptic flow vs. $p_{\rm T}$ of charged hadrons at $|\eta| < 0.8$ 
in Pb+Pb collisions at $\sqrt{s} = 2.76$~TeV with centrality 20$-$25\%.
Solid circles and solid squares are $v_2\{2\}$ and $v_2\{LYZ\}$
from CMS \protect\cite{Chatrchyan:2012ta}, open circles and histogram 
are $v_2\{EP\}$ and $v_2(\Psi_{EP,2})$ for HYDJET++ events, respectively.
(b) The same as (a) but for the triangular flow $v_3(p_{\rm T})$.
}
\label{fig1}
\end{figure}

We consider Pb+Pb collisions at $\sqrt{s} = 2.76$~TeV. HYDJET++ 
describes the differential elliptic and triangular flow quite well, as
one can see in Fig.~\ref{fig1}(a,b) where both flow harmonics,
calculated for events with centrality $\sigma/\sigma_{geo} = 20 - 30\%$,
are compared with the CMS data \cite{Chatrchyan:2012ta}. Recall, that 
the model employs ideal hydrodynamics, therefore, the falloff of both 
$v_2(p_{\rm T})$ and $v_3(p_{\rm T})$ after a certain transverse momentum 
about 2.5-3~GeV/$c$ is due to the jet influence. Jets themselves do not 
carry anisotropic flow apart from the small anisotropy caused by the jet
quenching. Therefore, when hadrons produced in hard processes begin to 
dominate the particle spectrum at $p_{\rm T} \geq 3$~GeV/$c$, the 
magnitudes of both harmonics, $v_2(p_{\rm T})$ and $v_3(p_{\rm T})$, 
drop.

\begin{figure}
\begin{center}
\resizebox{0.5\textwidth}{!}{
\includegraphics{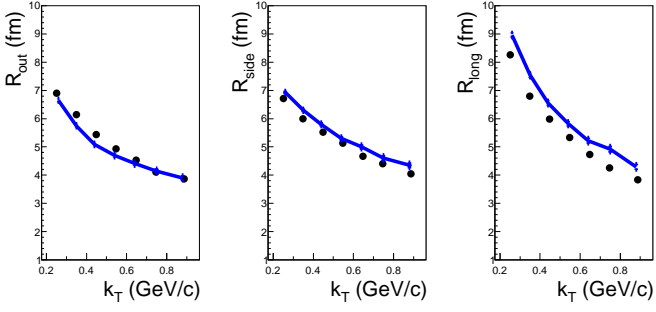}}
\end{center}
\caption{$\pi^\pm \pi^\pm$ correlation radii as functions of pion pair
transverse momentum $k_{\rm T}$ in 5\% of most central PbPb collisions 
at $\sqrt s_{\rm NN}=2.76$ TeV. Solid curves show the HYDJET++ 
calculations, full circles denote the ALICE data~\cite{Aamodt:2011mr}.} 
\label{fig2}
\end{figure}

Figure~\ref{fig2} shows the correlation radii $R_{out}$, $R_{side}$, 
and $R_{long}$ of charged pion pairs as functions of the pair transverse
momentum $k_{\rm T}$ with $|\eta|<0.8$ in 5\% of most central lead-lead 
collisions at $\sqrt s_{\rm NN}=2.76$ TeV. The results of the HYDJET++ 
simulation are plotted onto the ALICE data \cite{Aamodt:2011mr}. One can 
see that the model reproduces the measured $k_{\rm T}$-dependencies 
of the correlation radii $R_{out}$ and $R_{long}$ very well, and 
overestimates by about 8\% the corresponding distribution for $R_{long}$.

To describe both flow harmonics and femtoscopy observables 
simultaneously the whole set of parameters responsible for geometric and 
dynamical anisotropy was applied. Now, for the sake of clarity, we start
with investigation of individual influence of geometric and dynamical
deformations on elliptic flow and related to it second-order femtoscopy
radii oscillations. Recall briefly the observed main tendencies in the 
elliptic flow development and femtoscopy radii distributions with 
respect to the event plane $\Psi_{EP,2}$ \cite{Logg_npa14}. Differential
elliptic flow of charged particles, $v_2^{ch}(p_{\rm T})$ is positive, 
$R^2_{side}$ has a dip at $\Delta \phi_2 = \phi_{pair} - \Psi_{EP,2} 
\simeq \pi/2$, whereas $R^2_{out}$ demonstrates quite distinct maximum 
there. In contrast, distribution $R^2_{long}(\Delta \phi_2)$ appears 
to be rather flat. 

\begin{figure}
  \begin{center}
\resizebox{0.5\textwidth}{!}{%
   \includegraphics{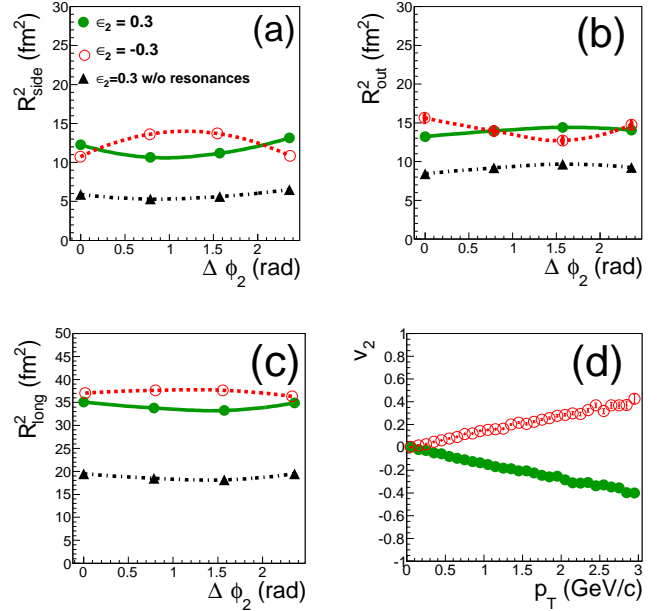}}
\caption{The azimuthal dependence of (a) $R^{2}_{side}$, 
(b) $R^{2}_{out}$, (c) $R^{2}_{long}$ as a function of 
$\Delta\phi_2 = \phi_{pair} - \Psi_{EP,2}$ for the centrality 20--30\% 
and $k_{\rm T}$ range 0.2--2.0~GeV/$c$, and (d) the azimuthal asymmetry 
coefficient $v_2$ versus $p_{\rm T}$.
Calculations were performed with the HYDJET++ model using the sets of 
parameters $\varepsilon_2$=0.3, $\delta_2$=0, $\varepsilon_3$=0, 
$\rho_3$=0 for directly produced particles (solid triangles) and 
for all particles after the resonance decays (solid circles); and 
$\varepsilon_2 = - 0.3$, $\delta_2$=0, $\varepsilon_3$=0, $\rho_3$=0 for
all particles (open circles). Lines are drawn to guide the eye.}
  \end{center}
\label{fig3:2_eps}
\end{figure}

The triangular flow is excluded in HYDJET++ by setting both 
triangularity parameters, $\varepsilon_3$ and $\rho_3$, to zero. Then, 
we consider isotropic expansion model in which the direction of the flow 
vector of a fluid cell coincides with its velocity vector. In this 
particular case $\delta_2 = 0$ and the only parameter causing the elliptic 
anisotropy of particle spectra is $\varepsilon_2$. 
Figure~\ref{fig3:2_eps} displays the azimuth distributions of the three 
radii squared together with differential elliptic flow $v_2(p_{\rm T})$ 
for calculations with $\varepsilon_2 = 0.3$ and $\varepsilon_2 = - 0.3$ 
of Pb+Pb collisions at centrality $20\% \leq \sigma/\sigma_{geo} \leq 
30\%$.
\footnote{From here the choice of the anisotropy parameters in the
model is arbitrary in order to demonstrate qualitative features of the
distributions. The statistics of generated events varies between
1.5 and 2 million Pb+Pb collisions with centrality $\sigma/\sigma_{geo} 
= 20 - 30\%$.}

We see that for $\varepsilon_2 = 0.3$ the radii squared reproduce 
qualitatively the trends observed experimentally, whereas the 
differential elliptic flow is negative at $p_{\rm T} \leq 5$~GeV/$c$. 
The latter result is obviously wrong. On the other hand, one can get 
positive $v_2(p_{\rm T})$ distribution by switching to negative value 
$\varepsilon_2 = - 0.3$, but in this case the azimuthal oscillations of 
$R^2_{side}$ and $R^2_{out}$ are out-of-phase, as depicted in 
Fig.~\ref{fig3:2_eps}. Decays of resonances increase all three 
radii but do not shift the phases of radii oscillations. Therefore, bare 
geometric anisotropy of the fireball cannot describe the true behavior of 
both elliptic flow and femtoscopic radii simultaneously.

\begin{figure}
\begin{center}
\resizebox{0.5\textwidth}{!}{%
    \includegraphics{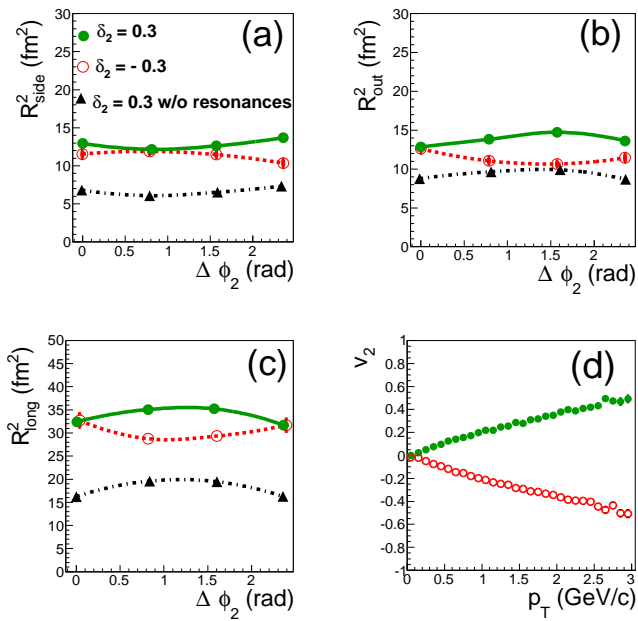}}
\caption{The same as Fig.~\ref{fig3:2_eps} but with $\varepsilon_2 = 0$, 
$\delta_2 = 0.3$ for particle spectra before (solid triangles) and 
after (solid circles) the resonance decays; and $\varepsilon_2 = 0$, 
$\delta_2 = - 0.3$ for all particles (open circles).}
\label{fig4:2_delta}
\end{center}
\end{figure}

If, in contrast, we will allow for only dynamical elliptic anisotropy 
in the system by setting $\varepsilon_2 = 0$ and $\delta_2 = \pm 0.3$, 
the picture will be drastically changed, as presented in 
Fig.~\ref{fig4:2_delta}. Positive value of $\delta_2$ provides positive
$p_{\rm T}$-differential elliptic flow, as well as local minimum of 
$R_{side}^2$ accompanied by local maximum of $R_{out}^2$ at $\Delta
\phi_2 \approx \pi/2$, respectively. Similar to the case with pure 
spatial anisotropy, the positions of extrema in radii oscillations are 
insensitive to the decays of resonances, although the magnitudes of the 
oscillations are increased. Calculations with negative value 
of $\delta_2$ result in a completely wrong behavior, namely, in negative 
elliptic flow and $\pi/2$-shift of the oscillation phase for the 
femtoscopy radii. 
 
Oscillations of the femtoscopic radii squared $R_{out}^2, R_{side}^2$,
and $R_{long}^2$ of charged pion pairs with respect to the triangular 
flow plane $\Psi_{EP,3}$ in Pb+Pb collisions at $\sqrt{s} = 2.76$~TeV 
were studied by ALICE Collaboration in \cite{alice_osc_psi3}. Several 
centralities from 10$-$20\% to 40$-$50\% were examined, and the general 
tendencies appeared to be as follows. In the interval $0 \leq \Delta
\phi_3 \leq 2,\ \Delta \phi_3 = \phi_{pair} - \Psi_{EP,3}$, both, 
$R_{side}^2$ and $R_{out}^2$, have maxima at $\pi/3$, whereas the 
distribution $R_{long}^2(\Delta \phi_3)$ is more flat, although the 
possible oscillations are not ruled out. It is interesting to note, that 
the signs and positions of the extrema in $R_\mu^2(\Delta \phi_2), \
\mu = {out, side}$, distributions in Pb+Pb at LHC energy exactly match 
those in Au+Au collisions at RHIC ($\sqrt{s} = 200$~GeV) 
\cite{phenix_npa13}. The oscillations of $R_\mu^2$ in the triangular
flow plane are more curious. Here $R_{out}^2$ reaches maximum at 
$\Delta \phi_3 = \pi/3$ in heavy-ion collisions at both RHIC and 
LHC energy. Distribution $R_{side}^2 (\Delta \phi_3)$, however, 
demonstrates maximum at $\Delta \phi_3 = \pi/3$ at LHC energy 
\cite{alice_osc_psi3}, and minimum at RHIC energy \cite{phenix_npa13}.

\begin{figure}
\begin{center}
\resizebox{0.5\textwidth}{!}{%
    \includegraphics{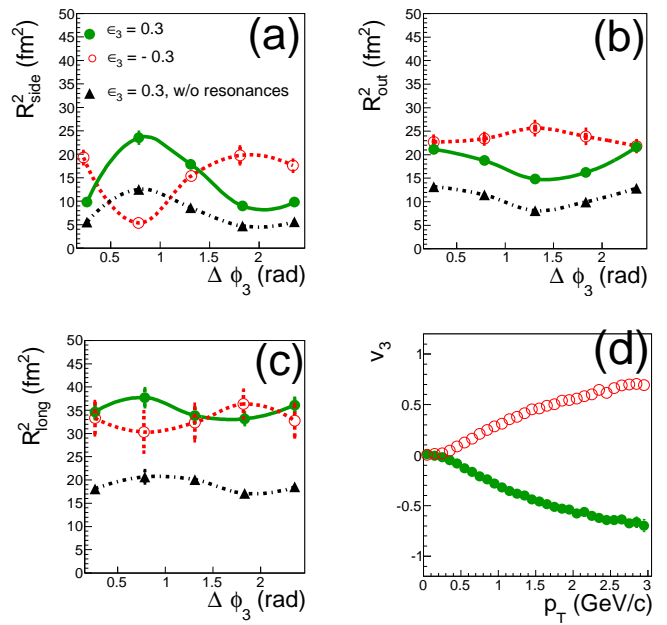}}
\caption{The azimuthal dependence of (a) $R^{2}_{side}$, 
(b) $R^{2}_{out}$, (c) $R^{2}_{long}$ as a function of 
$\Delta\phi_3 = \phi_{pair} - \Psi_{EP,3}$ for the centrality 20--30\% and 
$k_{\rm T}$ range 0.2--2.0~GeV/$c$, and (d) the azimuthal asymmetry 
coefficient $v_3$ versus $p_{\rm T}$.
Calculations were performed with the HYDJET++ model using the sets of 
parameters $\varepsilon_2$=0, $\delta_2$=0, $\varepsilon_3$=0.3, 
$\rho_3$=0 for directly produced particles (solid triangles) and for all 
particles after the resonance decays (solid circles); and 
$\varepsilon_2$=0, $\delta_2$=0, $\varepsilon_3 = -0.3$, $\rho_3$=0 for 
all particles (open circles). Lines are drawn to guide the eye.} 
\label{fig5:3_eps}
\end{center}
\end{figure}

Let us investigate the oscillations of femtoscopic radii in geometry 
dominated scenario, where only the third harmonic of anisotropic flow
is present. This means that we set to zero $\varepsilon_2,\ \delta_2$, 
and $\rho_3$, and employ $\varepsilon_3 \neq 0$ as the only parameter
responsible for the triangular flow generation. Two opposite cases, 
$\varepsilon_3 = -0.3$ and $\varepsilon_3 = 0.3$, are presented in 
Fig.~\ref{fig5:3_eps}. For negative value of the spatial triangularity,
$R_{side}^2$ has a distinct minimum at $\Delta \phi_3 \simeq 0.75~rad$ 
and smeared maximum at $\Delta \phi_3 \simeq 1.75~rad$, $R_{out}^2$ has 
a maximum at $\Delta \phi_3 \simeq 1.3~rad$, $R_{long}^2$ is almost
independent on $\Delta \phi_3$, and the positive differential 
triangular flow $v_3(p_{\rm T})$ increases with rising transverse 
momentum. For positive spatial triangularity, the behaviour of 
$R_\mu^2$ and $v_3(p_{\rm T})$ is completely opposite: $R_{side}^2$ has 
maximum at $\Delta \phi_3 \approx 0.75~rad$ and minimum at $\Delta 
\phi_3 \approx 1.75~rad$, $R_{out}^2$ reaches minimum at 
$\Delta \phi_3 \approx 1.3~rad$, the distribution $R_{long}^2(\Delta 
\phi_3)$ is flat within the error bars. Negative for all transverse 
momenta $v_3(p_{\rm T})$ drops with increasing $p_{\rm T}$. 
Again, decays of resonances increase the femtoscopic radii and 
magnitudes of their oscillations. However, the positions of extrema of 
the $R^2_\mu (\Delta \phi_3),\ \mu = out,side,long$ spectra of directly 
produced hadrons stay put. Therefore, neither of two 
scenarios, (i) with positive or (ii) with negative $\varepsilon_3$, is 
fully consistent qualitatively with the experimentally observed signals.
                
\begin{figure}
\begin{center}
\resizebox{0.5\textwidth}{!}{%
    \includegraphics{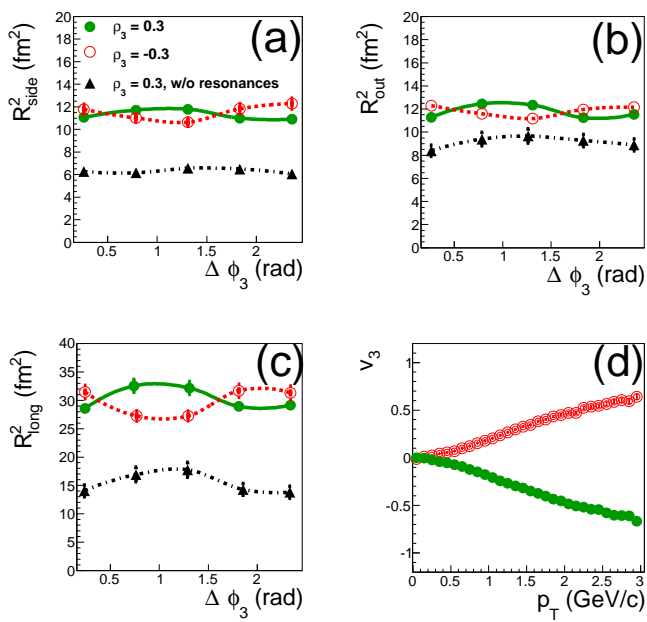}}
\caption{The same as Fig.~\ref{fig5:3_eps} but with $\varepsilon_3=0$, 
$\rho_3 = 0.3$ for particle spectra before (solid triangles) and
after (solid circles) the resonance decays; and $\varepsilon_3 = 0$, 
$\rho_3 = -0.3$ for all particles (open circles).}
\label{fig6:3_rho}
\end{center}
\end{figure}

In the scenario with dynamical triangular anisotropy domination, one
sets $\varepsilon_2 = \varepsilon_3 = \delta_2 = 0$, and $\rho_3 \neq 0$. 
Again, $\rho_3$ can be positive, e.g. $\rho_3 = 0.3$, and negative, 
$\rho_3 = - 0.3$. Calculations with both sets of parameters are shown in 
Fig.~\ref{fig6:3_rho}. For positive value of $\rho_3$ the experimentally
observed behaviour of femtoscopic radii and triangular flow is 
quantitatively reproduced. Namely, both $R^{2}_{out}$ and
$R^{2}_{side}$ have not very distinct maxima at $\Delta \phi_3 
\simeq \pi/3$, and differential triangular flow $v_3(p_{\rm T})$ is
positive. $R^{2}_{long}$ oscillates slightly, but linear fit within 
the error bars is still possible. The phases of the oscillations 
are not shifted after the decays of resonances.

For the case with negative value of the $\rho_3$ all three femtoscopic 
radii squared, $R_\mu^2(\Delta \phi_3),\ \mu = out, side, long$, 
demonstrate the out-of-phase behaviour, compared to the case with
positive $\rho_3$, see Fig.~\ref{fig6:3_rho}. Differential triangular 
flow is also negative. The dynamical anisotropy alone, however, cannot 
reproduce the magnitudes of the oscillations. Therefore, the final 
signal appears as a superposition of dynamical anisotropy and 
geometrical anisotropy.

\begin{figure}
\begin{center}
\resizebox{0.50\textwidth}{!}{%
    \includegraphics{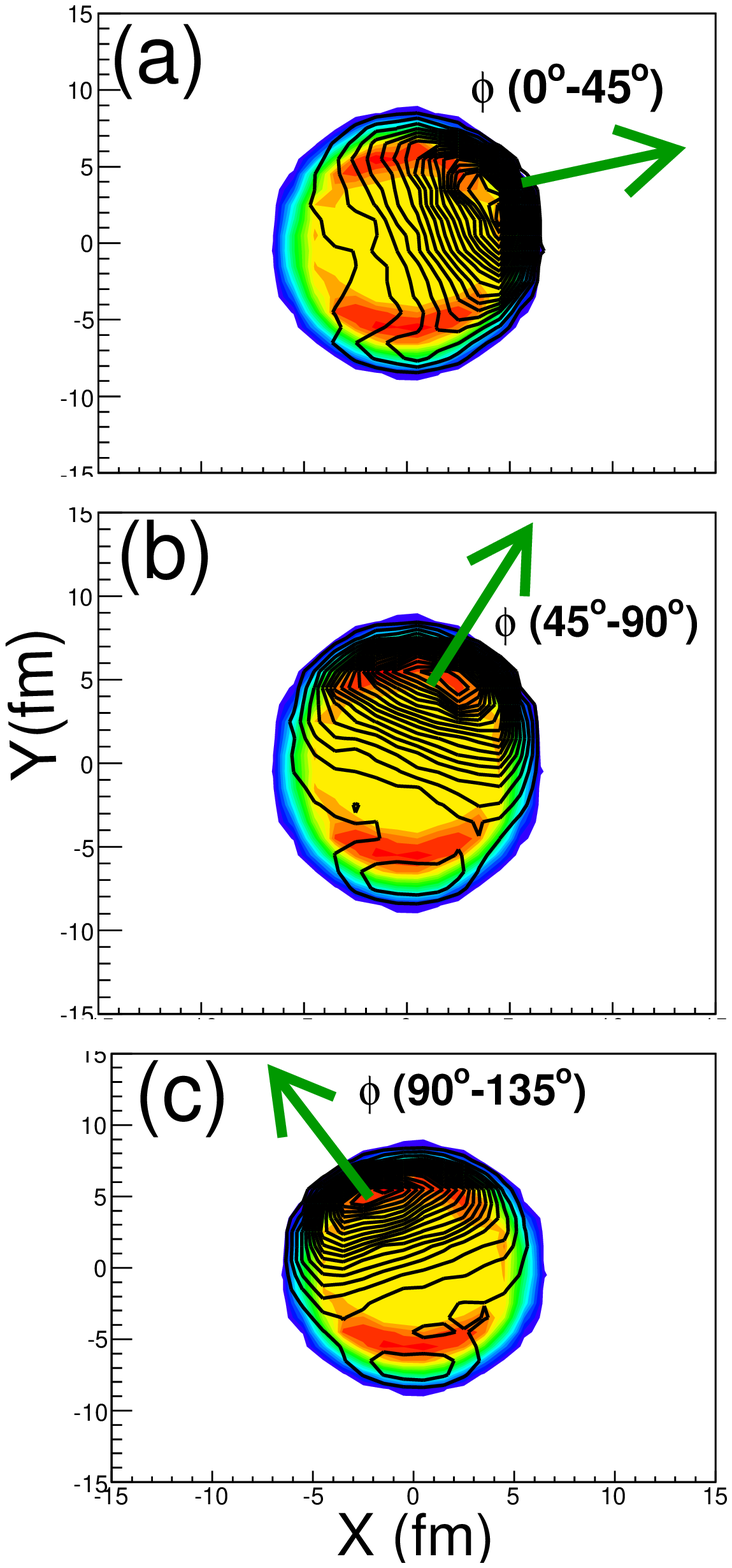}
\hspace*{-1.5cm}
    \includegraphics{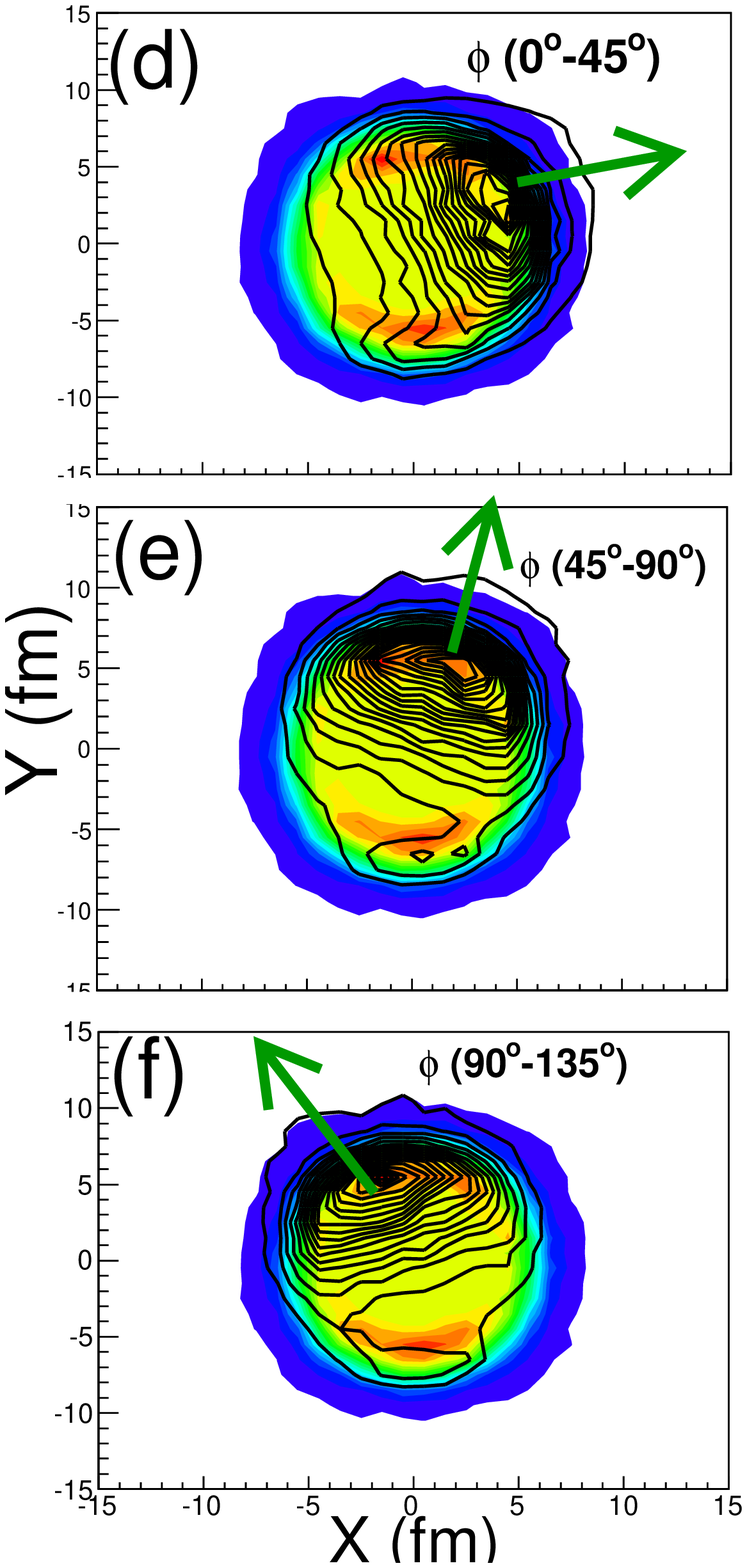}}
\caption{
(Color online)
Pion emission function before (left column) and after (right column) 
decays of resonances in the transverse plane of HYDJET++ simulated
Pb+Pb collisions at $\sqrt{s} = 2.76$~TeV with centrality 20--30\%.
Only spatial elliptic anisotropy with $\varepsilon_2 = 0.5$ is 
considered, while the remaining anisotropy parameters $\delta_2,
\varepsilon_3,$ and $\rho_3$ are taken to be zero. Shaded contours are 
identical for each column and indicate the density of emitted pions.
Contour lines show the densities of pions emitted at angles $0 < \phi
\leq \pi/4$ (upper row), $\pi/4 < \phi \leq \pi/2$ (middle row), and 
$\pi/2 < \phi \leq 3\pi/4$ (bottom row), respectively. 
}
\label{fig7:2_geo}
\end{center}
\end{figure}

\begin{figure}
\begin{center}
\resizebox{0.5\textwidth}{!}{%
    \includegraphics{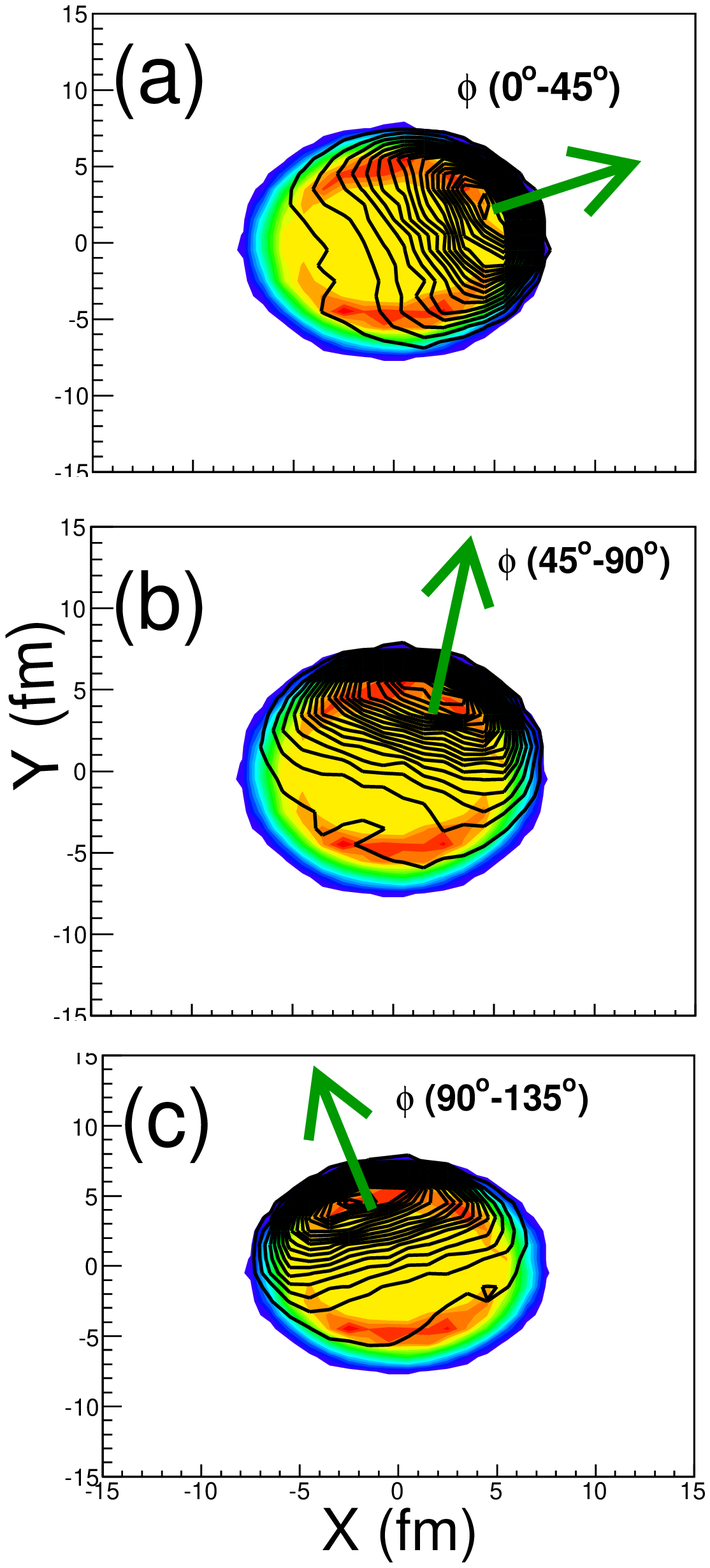}
\hspace*{-1.5cm}
    \includegraphics{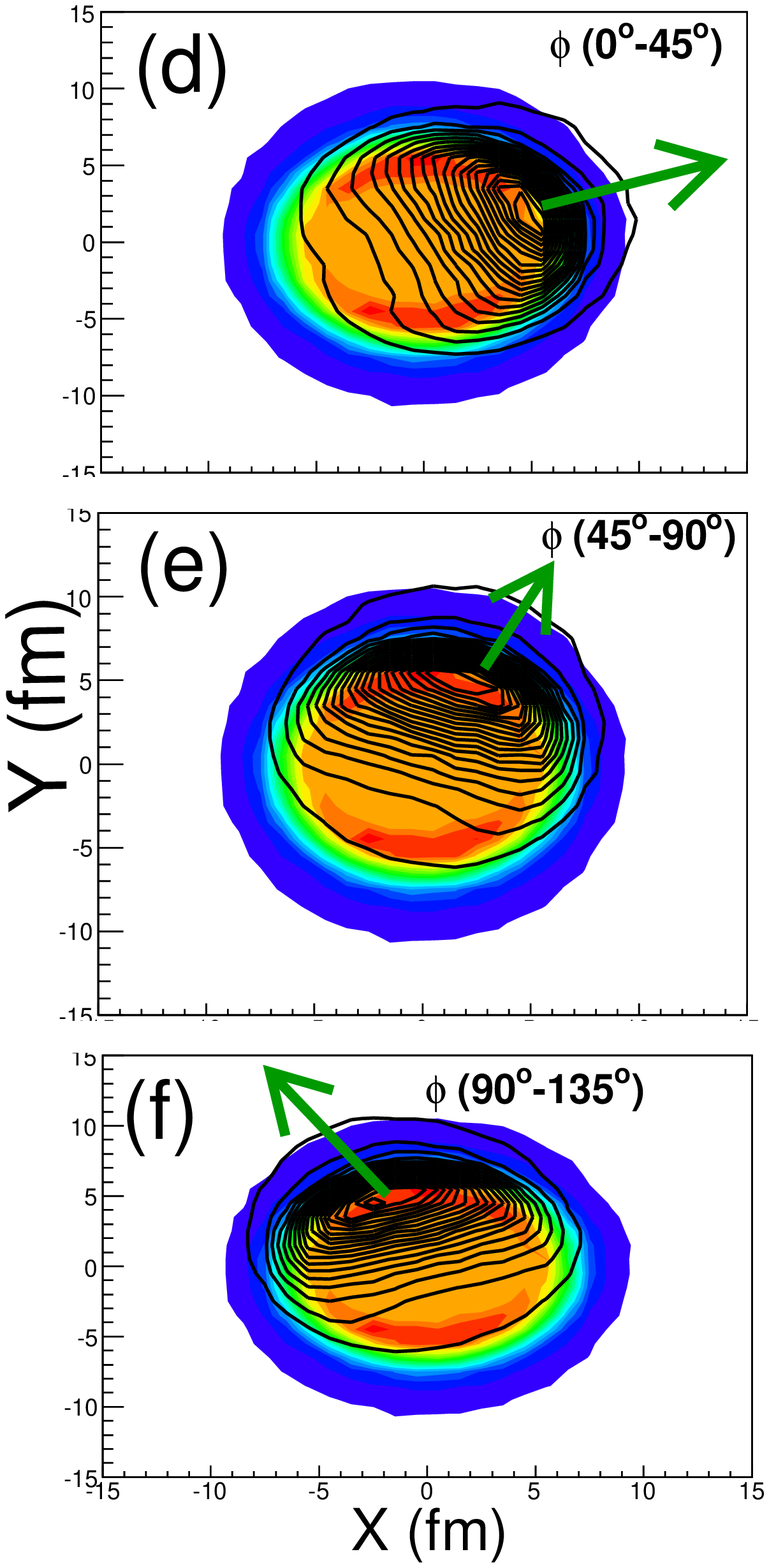}}
\caption{
(Color online)
The same as Fig.~\ref{fig7:2_geo} but for calculations with non-zero
dynamical elliptic anisotropy, $\delta_2 = -0.3$, while other anisotropy
parameters are equal to zero, $\{\varepsilon_2, \varepsilon_3, \rho_3 
\} = 0$. }
\label{fig8:2_dyn}
\end{center}
\end{figure}

\begin{figure}
\begin{center}
\resizebox{0.5\textwidth}{!}{%
    \includegraphics{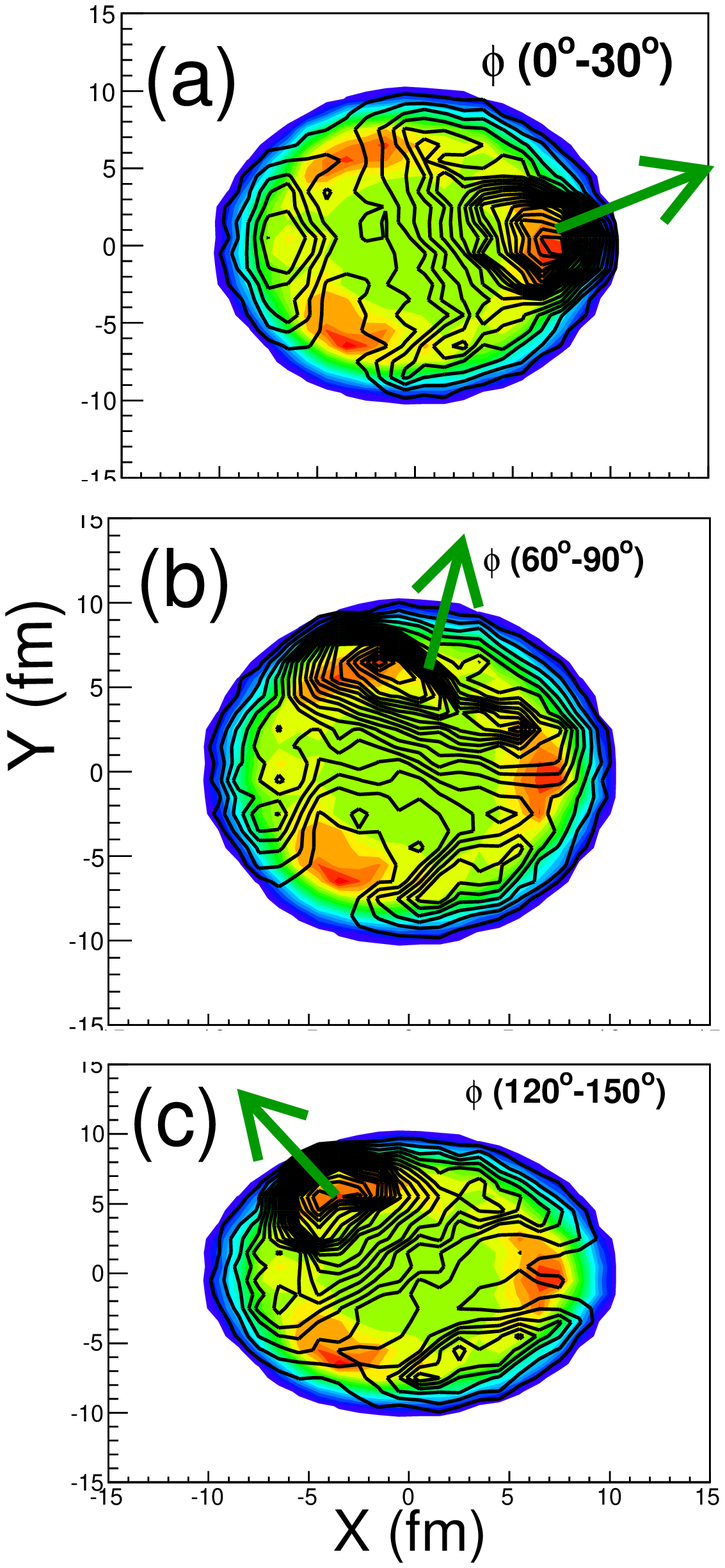}
\hspace*{-1.5cm}
    \includegraphics{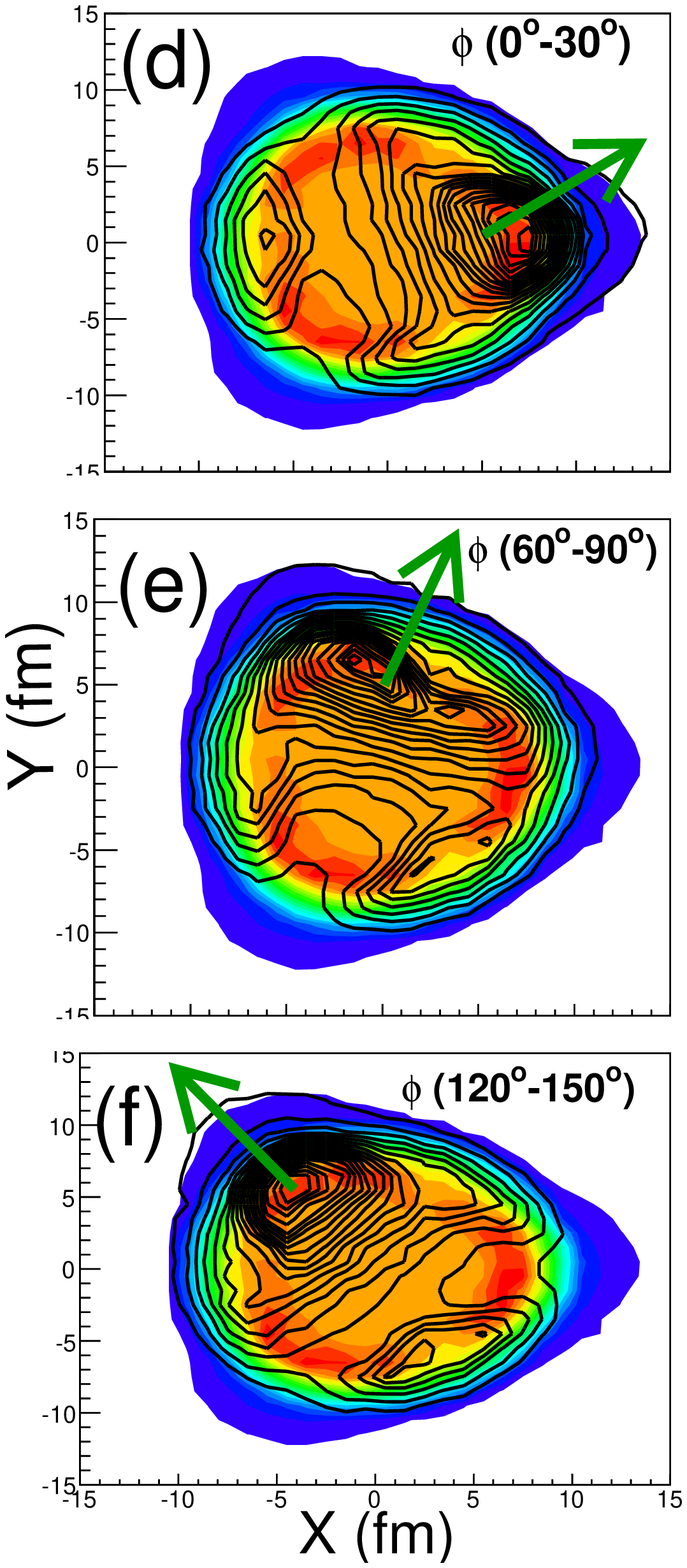}}
\caption{
(Color online)
The same as Fig.~\ref{fig7:2_geo} but for calculations with non-zero
geometric triangular anisotropy, $\varepsilon_3 = 0.3$, while other 
anisotropy parameters are equal to zero, $\{\varepsilon_2, \delta_2, 
\rho_3 \} = 0$.
Contour lines show the densities of pions emitted at angles $0 < \phi
\leq \pi/6$ (upper row), $\pi/3 < \phi \leq \pi/2$ (middle row), and 
$2\pi/3 < \phi \leq 5\pi/6$ (bottom row), respectively. } 
\label{fig9:3_geo}
\end{center}
\end{figure}

\begin{figure}
\begin{center}
\resizebox{0.5\textwidth}{!}{%
    \includegraphics{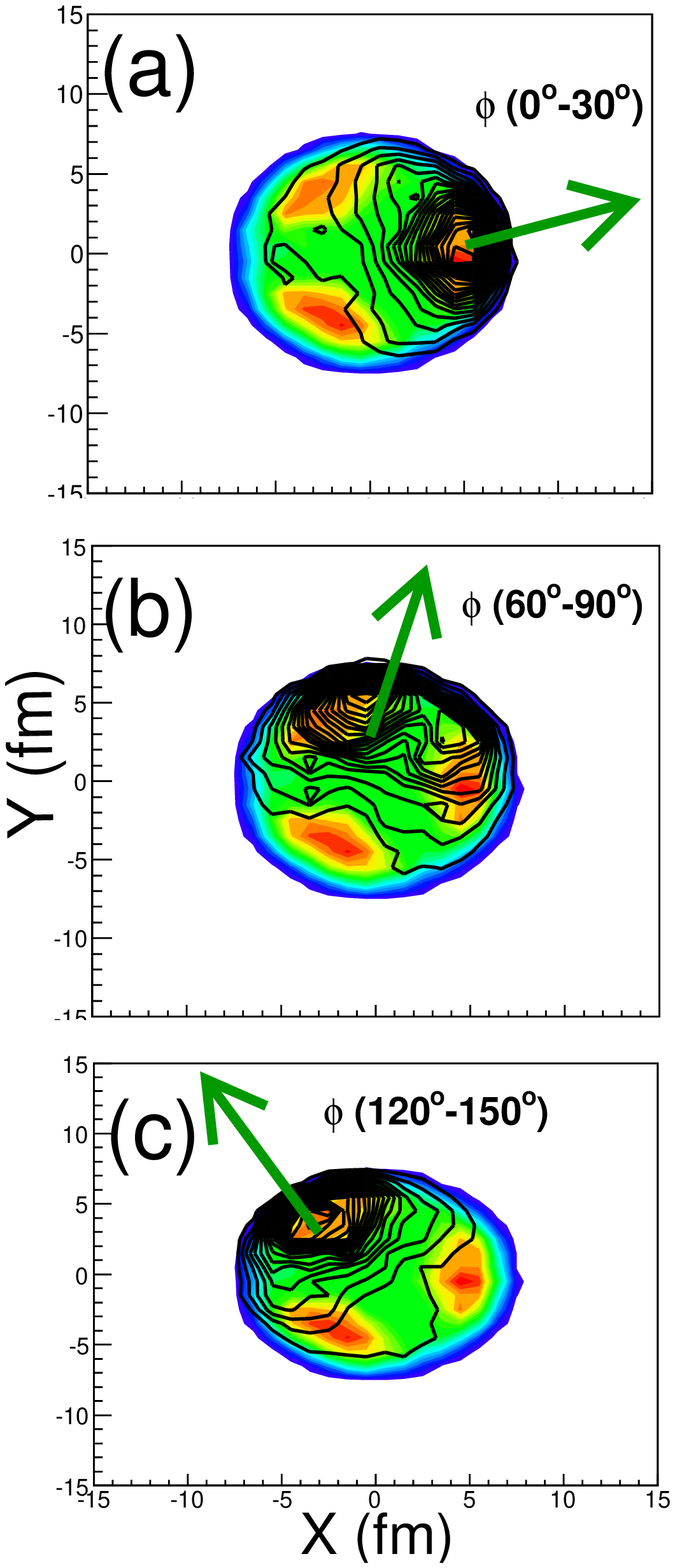}
\hspace*{-1.5cm}
    \includegraphics{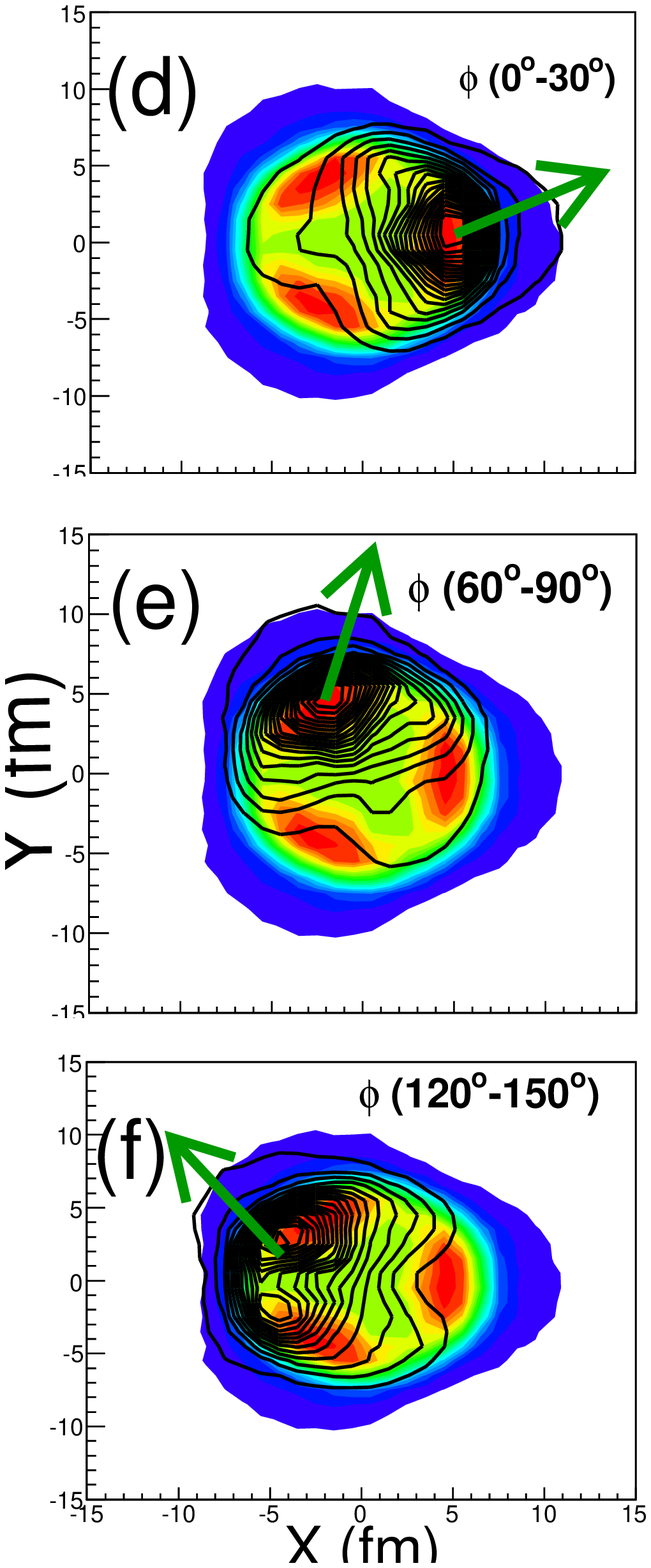}}
\caption{
(Color online)
The same as Fig.~\ref{fig7:2_geo} but for calculations with non-zero
dynamical triangular anisotropy, $\rho_3 = 0.5$, while other anisotropy
parameters are taken to be zero, $\{\varepsilon_2, \delta_2, 
\varepsilon_3 \} = 0$. }
\label{fig10:3_dyn}
\end{center}
\end{figure}

The femtoscopy analysis allows us to probe the emission zones, also 
known as ```homogeneity lengths", rather than the sizes of the whole
source \cite{led04}. To study the size and shape of the particle sources
in HYDJET++, we plot in Figs.~\ref{fig7:2_geo}$\div$\ref{fig10:3_dyn} the
transverse plane emission densities of pions, radiated directly from the 
freeze-out hypersurface (left columns), and of all pions, produced both
at the freeze-out hypersurface and from the resonance decays (right 
columns). Figure~\ref{fig7:2_geo} corresponds to the presence of only
geometric ellipticity, $\varepsilon_2 = 0.5$, while other dynamic and 
spatial sources of the system ellipticity and triangularity are absent,
i.e. $\delta_2 = \varepsilon_3 = \rho_3 =0$. Other figures illustrate
the cases with $\delta_2 = -0.3$ and $\varepsilon_2 = \varepsilon_3 = 
\rho_3 =0$ (Fig.~\ref{fig8:2_dyn}), with $\varepsilon_3 = 0.3$ and
$\varepsilon_2 = \delta_2 = \rho_3 =0$ (Fig.~\ref{fig9:3_geo}), and
with $\rho_3 =0.5$ and $\varepsilon_2 = \delta_2 = \varepsilon_3 = 0$
(Fig.~\ref{fig10:3_dyn}), respectively. Three angular areas of the pion 
emission are considered for each of the anisotropies. For the bare 
elliptic anisotropy we opted for (i) $0 < \phi \leq \pi/4$, (ii) $\pi/4 
<\phi \leq \pi/2$, and (iii) $\pi/2 < \phi \leq 3\pi/4$, shown in 
Fig.~\ref{fig7:2_geo} and Fig.~\ref{fig8:2_dyn}, whereas for the 
triangular anisotropy the choice is (i) $0 < \phi \leq \pi/6$, (ii)
$\pi/3 < \phi \leq \pi/2$, and (iii) $2\pi/3 < \phi \leq 5\pi/6$,
displayed in Fig.~\ref{fig9:3_geo} and Fig.~\ref{fig10:3_dyn}. Let us
compare first emission functions corresponding to geometric and 
dynamical elliptic anisotropies, presented in Fig.~\ref{fig7:2_geo} and
Fig.~\ref{fig8:2_dyn}. One can see that the same-density contours of 
pion emission are much smoother for dynamical anisotropy compared to the 
spatial one. Also, the spatial anisotropy provides stronger difference 
between the emitting zones at three different angles in contrast to the
dynamical anisotropy. This circumstance explains the stronger angular
dependence of $R^2_{out}(\Delta \phi_2)$ and $R^2_{side}(\Delta \phi_2)$
for geometric anisotropy seen in Fig.~\ref{fig3:2_eps}. Similar 
difference between the spatial and dynamical anisotropy was also found 
recently in the Buda-Lund model in \cite{CTCL_17}. Pions coming from the
decays of resonances enlarge the emission areas and make the density 
contours smoother, as seen in right windows of both Fig.~\ref{fig7:2_geo}
and Fig.~\ref{fig8:2_dyn}.          

The pion emission functions for the systems with only geometric or only 
dynamical triangularity, depicted in Fig.~\ref{fig9:3_geo} and in
Fig.~\ref{fig10:3_dyn}, respectively, reveal similar tendencies. First 
of all, the spatial triangularity produces much stronger difference for 
the emitting areas in three azimuthal directions compared to the 
dynamical triangularity. Then, as in the elliptic anisotropy case, 
decays of resonances lead to rounding of the pion density contours. The 
triangular profile of the outher area of transverse pion emission 
becomes quite distinct after the resonance decays. All four systems 
shown in Figs.~\ref{fig7:2_geo}$\div$\ref{fig10:3_dyn} demonstrate a 
clear angular dependence of the homogeneity regions discussed, e.g., in 
\cite{Wiedemann:1997cr,H_prl99} in addition to that given by the set of 
Eqs.(\ref{eq:radii}). It is worth noting also that the non-Gaussian 
shapes of the homogeneity regions make complicated restoration of the 
shape and size of the source by the standard femtoscopic analysis.       

\section{Conclusions}
\label{concl}

Second- and third-order oscillations of the femtoscopic radii 
$R_{side}^2, R_{out}^2$, and $R_{long}^2$ in Pb+Pb collisions at 
$\sqrt{s} = 2.76$~TeV were studied within the HYDJET++ model together
with the differential elliptic and triangular flow. For each type of 
the flow harmonics the model assumes two parameters. One of them is 
responsible for spatial, or geometric, deformation of the freeze-out
hypersurface, whereas the other parameter is accountable for the 
dynamical flow anisotropy, respectively. By switching on and off of 
these key parameters one can investigate the influence of separated 
spatial and dynamical anisotropy effects on the flow harmonics and on 
the femtoscopic radii. Our study indicates that merely geometric 
anisotropy cannot reproduce simultaneously the correct phase of the 
second- and third-order oscillations of the femtoscopic radii and the
corresponding differential flow harmonics. Dynamical flow anisotropy,
in contrast, provides correct qualitative description of both 
$p_{\rm T}$-dependence of the flow harmonics and the phases of the 
femtoscopic radii oscillations. 

The spatial anisotropy, however, reveals stronger difference between 
the emitting zones of pions, radiated at different angles. This leads to
stronger azimuthal oscillations of femtoscopic radii in case with 
spatial ellipticity or triangularity compared to the case with the 
dynamical ones. Our findings are in line with the results of other 
models \cite{PSH_13,CTCL_17}. 
Decays of resonances provide significant increase of the emitting areas 
in the both planes of elliptic and triangular flow. Also, the resonance 
decays make the radii oscillations more pronounced, but they do not 
change the phases of the oscillations.
For the quantitative description of the flow and the femtoscopy
observables one has to use the full set of geometric and dynamical
anisotropy parameters.

\begin{acknowledgement}
Fruitful discussions with R. Lednicky are gratefully acknowledged.
This work was supported in parts by the grant from the President of 
Russian Federation for Scientific Schools (Grant No. 7989.2016.2); and
the Norwegian Research Council (NFR) under grant No. 255253/F50 - CERN 
Heavy Ion Theory. 
\end{acknowledgement}



\begin{thebibliography}{}


\bibitem{Shur_04}
E.~Shuryak, Prog. Part. Nucl. Phys. {\bf 53}, 273 (2004)

\bibitem{Land_53} L.D.~Landau,
Izv. Akad. Nauk SSSR, Ser. Fiz. {\bf 17}, 51 (1953) (in Russian);
S.Z.~Belenkij and L.D.~Landau,
Nuovo Cimento Suppl. {\bf 3}, 15 (1956)

\bibitem{Bjor_83} 
J.D.~Bjorken, Phys. Rev. D {\bf 27}, 140 (1983)

\bibitem{VoZh_96} 
S.A.~Voloshin, Y.~Zhang, Z. Phys. C {\bf 70}, 665 (1996)

\bibitem{VPS_10}
S.A.~Voloshin, A.M.~Poskanzer, R.~Snellings,
in {\it Relativistic Heavy Ion Physics\/}, Landolt-B{\"o}rnstein
Database Vol. {\bf 23}, edited by R.~Stock (Springer, Berlin, 2010), 
p.5$-$54.

\bibitem{HBT}
R.~Hanbury~Brown, R.Q.~Twiss, Phil. Mag. Ser.7 {\bf 45}, 663 (1954) 

\bibitem{GGL} G.~Goldhaber, S.~Goldhaber, W.-Y.~Lee, A.~Pais,
Phys. Rev. {\bf 120}, 300 (1960)

\bibitem{pod89} M.I.~Podgoretsky,
Fiz. Elem. Chast. Atom. Yadra {\bf 20}, 628 (1989) (in Russian)

\bibitem{led04} R.~Lednicky, Phys. Atom. Nucl. {\bf 67}, 72 (2004)

\bibitem{lis05} M.~Lisa, S.~Pratt, R.~Soltz, U.~ Wiedemann,
Ann. Rev. Nucl. Part. Sci. {\bf 55}, 357 (2005)

\bibitem{pod83} M.I.~Podgoretsky,
Sov. J. Nucl. Phys. {\bf 37}, 272 (1983); \\
R.~Lednicky, {\it preprint\/} JINR B2-3-11460, (Dubna, 1978); \\
P.~Grassberger, Nucl. Phys. B {\bf 120}, 231 (1977)

\bibitem{bdh94}
G.F.~Bertsch, P.~Danielewicz, M.~Herrmann,
Phys. Rev. C {\bf 49}, 442 (1994); \\
S.~Pratt, in {\it Quark Gluon Plasma 2\/}, edited by R.C.~Hwa,
(World Scientific, Singapore, 1995), p.700; \\
S.~Chapman, P.~Scotto, U.~Heinz,
Phys. Rev. Lett. {\bf 74}, 4400 (1995)

\bibitem{Wiedemann:1997cr} 
U.~Wiedemann, Phys. Rev. С {\bf 57}, 266 (1998)

\bibitem{LHW_plb00} 
M.A.~Lisa, U.~Heinz, U.A.~Wiedemann, Phys. Lett. B {\bf 489}, 287 (2000)

\bibitem{HK_plb02} 
U.~Heinz, P.F.~Kolb, Phys. Lett. B {\bf 542}, 216 (2002)

\bibitem{RL_prc04}
F.~Retiere, M.A.~Lisa, Phys. Rev. C {\bf 70}, 044907 (2004)

\bibitem{CTC_epja08}
M.~Csanad, B.~Tomasik, T.~Csorgo, Eur. Phys. J. A {\bf 37}, 111 (2008)

\bibitem{PSH_13}
C.J.~Plumberg, C.~Shen, U.~Heinz, Phys. Rev. C {\bf 88}, 044914 (2013)

\bibitem{LCTC_epja16}
S.~L{\"o}k{\"o}s, M.~Csanad, B.~Tomasik, T.~Csorgo,
Eur. Phys. J. A {\bf 52}, 311 (2016)

\bibitem{CTCL_17}
J.~Cimerman, B.~Tomasik, M.~Csanad, S.~L{\"o}k{\"o}s, 
Eur. Phys. J. A {\bf 53}, 161 (2017) 

\bibitem{CL_prc96} 
T.~Csorgo, B.~Lorstad, Phys. Rev. C {\bf 54}, 1390 (1996)

\bibitem{SR_prl79} 
P.J.~Siemens, J.O.~Rasmussen, Phys. Rev. Lett {\bf 42}, 880 (1979)

\bibitem{Lokhtin:2008xi} I.P.~Lokhtin, L.V~Malinina, S.V.~Petrushanko,
A.M.~Snigirev, I.~Arsene, K.~Tywoniuk,
Comput. Phys. Commun. {\bf 180}, 779 (2009)

\bibitem{Lokhtin:2012re} I.P.~Lokhtin, A.V.~Belyaev, L.V~Malinina,
S.V.~Petrushanko, E.P.~Rogochaya, A.M.~Snigirev,
Eur. Phys. J. C {\bf 72}, 2045 (2012)

\bibitem{Bravina:2013xla} L.V.~Bravina, B.H.~Brusheim Johansson,
G.Kh.~Eyyubova, V.L.~Korotkikh,  I.P.~Lokhtin, L.V.~Malinina,
S.V.~Petrushanko, A.M.~Snigirev, E.E.~Zabrodin,
Eur. Phys. J. C {\bf 74}, 2807 (2014)

\bibitem{THERM}
M.~Chojnacki, A.~Kisiel, W.~Florkowski, W.~Broniowski,
Comput. Phys. Commun. {\bf 183}, 746 (2012)


\bibitem{Amelin:2006qe} N.S.~Amelin et al., 
Phys. Rev. {\bf C 74}, 064901 (2006)

\bibitem{Amelin:2007ic} N.S.~Amelin et al., 
Phys. Rev. {\bf C 77}, 014903 (2008)

\bibitem{Lokhtin:2005px} I.P.~Lokhtin, A.M.~Snigirev, 
Eur. Phys. J. C {\bf 45}, 211 (2006)

\bibitem{Sjostrand:2006za} T.~Sjostrand, S.~Mrenna, P.~Skands, 
JHEP {\bf 0605}, 026 (2006)

\bibitem{Baier:1996kr}  
R.~Baier, Yu.L.~Dokshitzer, A.H.~Mueller, S.~Peigne, D.~Schiff,
Nucl. Phys. B {\bf 483}, 291 (1997)

\bibitem{Baier:1999ds}  
R.~Baier, Yu. L.~Dokshitzer, A.H.~Mueller, D.~Schiff,
Phys. Rev. {\bf C 60}, 064902 (1999)

\bibitem{Baier:2001qw} 
R.~Baier, Yu. L.~Dokshitzer, A.H.~Mueller, D.~Schiff,
Phys. Rev. C {\bf 64}, 057902 (2001)

\bibitem{Bjorken:1982tu} 
J.D.~Bjorken, Fermilab publication Pub-82/29-THY (1982)

\bibitem{Braaten:1991jj} E.~Braaten, M.~Thoma, 
Phys. Rev. D {\bf 44}, 1298 (1991)

\bibitem{Lokhtin:2000wm} I.P.~Lokhtin, A.M.~Snigirev, 
Eur. Phys. J. C {\bf 16}, 527 (2000)

\bibitem{Tywoniuk:2007xy} 
K.~Tywoniuk, I.C.~Arsene, L.~Bravina, A.B.~Kaidalov, E.~Zabrodin, 
Phys. Lett. B {\bf 657}, 170 (2007)

\bibitem{v2_prc09}
G.~Eyyubova, L.~Bravina, V.L.~Korotkih, I.P.~Lokhtin, L.V.~Malinina,
S.V.~Petrushanko, A.M.~Snigirev, E.~Zabrodin,
Phys. Rev. C {\bf 80}, 064907 (2009)

\bibitem{v2_sqm09}
E.E.~Zabrodin, L.V.~Bravina, G.Kh.~Eyyubova, I.P.~Lokhtin, L.V.~Malinina,
S.V.~Petrushanko, A.M.~Snigirev,
J. Phys. G {\bf 37}, 094060 (2010)

\bibitem{v3_prc17}
J.~Crkovsk\'{a} et al.,
Phys. Rev. C {\bf 95}, 014910 (2017)

\bibitem{v3_sqm15}
E.E.~Zabrodin, L.V.~Bravina, B.H.~Brusheim Johansson, J.~Crkovsk\'{a},
G.Kh.~Eyyubova, V.L.~Korotkikh, I.P.~Lokhtin, L.V.~Malinina,
S.V.~Petrushanko, A.M. Snigirev,
J. Phys.: Conf. Ser. {\bf 668}, 012099 (2016)

\bibitem{v4_prc13}
L.~Bravina, B.H.~Brusheim Johansson, G.~Eyyubova, E.~Zabrodin,
Phys. Rev. C {\bf 87}, 034901 (2013)

\bibitem{v6_prc14}
L.V.~Bravina, B.H.~Brusheim Johansson, G.Kh.~Eyyubova, V.L.~Korotkikh,
I.P.~Lokhtin, L.V.~Malinina, S.V.~Petrushanko, A.M.~Snigirev, 
E.E.~Zabrodin,
Phys. Rev. C {\bf 89}, 024909 (2014)

\bibitem{ridge_prc15}
G.Kh.~Eyyubova, V.L.~Korotkikh, I.P.~Lokhtin, S.V.~Petrushanko,
A.M.~Snigirev, L.V.~Bravina, E.E.~Zabrodin,
Phys. Rev. C {\bf 91}, 064907 (2015)

\bibitem{fluct_epjc15}
L.V.~Bravina, E.S.~Fotina, V.L.~Korotkikh, I.P.~Lokhtin, L.V.~Malinina,
E.N.~Nazarova, S.V.~Petrushanko, A.M.~Snigirev, E.E.~Zabrodin,
Eur. Phys. J. C {\bf 75}, 588 (2015)

\bibitem{charm_arx}
I.P.~Lokhtin, A.V.~Belyaev, G.~Ponimatkin, E.Yu.~Pronina, G.Kh.~Eyyubova, 
J. Exp. Theor. Phys. {\bf 124}, 244 (2017)




\bibitem{Chatrchyan:2012ta}
S.~Chatrchyan et al. (CMS Collaboration),  
Phys. Rev. C {\bf 87}, 014902 (2013)

\bibitem{Aamodt:2011mr}
K.~Aamodt et al. (ALICE Collaboration), 
Phys. Lett. B {\bf 696}, 328 (2011)

\bibitem{Logg_npa14}
V.~Loggins et al. (ALICE Collaboration), 
Nucl. Phys. A {\bf 931}, 1088 (2014)

\bibitem{alice_osc_psi3} 
M.~Saleh et al. (ALICE Collaboration), arXiv:1704.06206

\bibitem{phenix_npa13}
T.~Niida et al. (PHENIX Collaboration),
Nucl. Phys. A {\bf 904-905}, 439c (2013)

\bibitem{H_prl99} 
H.~Heiselberg, Phys. Rev. Lett. {\bf 82}, 2052 (1999)

\end{thebibliography}
\end{document}